\documentclass[%
 reprint,
superscriptaddress,
 amsmath,amssymb,
 aps,
floatfix,
]{revtex4-2}

\usepackage{amsmath,amssymb}
\usepackage[colorlinks=true,allcolors=blue]{hyperref}

\newcommand{\ket}[1]{\ensuremath{|#1\rangle}}
\newcommand{\CX}[2]{\mathrm{CNOT}_{#1\to #2}}
\newcommand{\MX}{M_X}
\newcommand{\MZ}{M_Z}

\begin{document}

\title{CSS circuits are ICM circuits without magic states}

\author{Alexandru Paler}
\affiliation{Aalto University, School of Science, Department of Computer Science, Konemiehentie~2, 02150 Espoo, Finland}

\begin{abstract}
Gidney and Jones recently introduced ``CSS circuits'', generated by $X$, $Z$, $\MX$, $\MZ$ and CNOT, and showed that they can be sampled with $O(1)$ work per gate. They suspected that the class had been seen before. We point out that it is, up to a normal form, the initialisation--CNOT--measurement (ICM) form introduced in 2015 for compiling fault-tolerant circuits, restricted to $\ket{0}$ and $\ket{+}$ initialisations. The ICM construction answers constructively their question about ``non-CSS-ness as a resource'': every Clifford$+T$ circuit compiles to a CSS circuit acting on $\ket{Y}$ and $\ket{A}$ input states.
\end{abstract}

\maketitle

\textit{CSS circuits.}---Ref.~\cite{GJ26} defines a CSS circuit as one built from $X$, $Z$, $\MX$, $\MZ$ and CNOT, applied possibly probabilistically and adaptively. Because such circuits are degenerate stabilizer circuits, the all-zero record is a valid reference sample and Pauli-frame simulation~\cite{Gid21} runs in $O(1)$ per gate, even under atom loss. The authors write that they suspect they have rediscovered the concept, pointing to phase-free ZX diagrams~\cite{Kis22}.

\textit{The ICM form.}---Refs.~\cite{PPND15,PPND17} introduced the ICM form as the target representation of a fault-tolerant compiler. An ICM circuit has (I) a layer of initialisations in $\ket{0}$, $\ket{+}$, $\ket{Y}=\ket{0}+i\ket{1}$ or $\ket{A}=\ket{0}+e^{i\pi/4}\ket{1}$; (C) a fixed array of CNOTs; (M) a time-ordered series of $\MX$ and $\MZ$ measurements, whose basis may depend on earlier outcomes. Pauli corrections are not applied but tracked in a Pauli frame~\cite{PDNP14}. The gate set was chosen precisely because $X/Z$ preparation and measurement and CNOT are the operations native to CSS codes, including the surface code.

\textit{Equivalence.}---Let $\mathcal{C}$ be a CSS circuit in which classical control acts only on Pauli gates and on the choice of measurement basis. Then $\mathcal{C}$ is equivalent to an ICM circuit whose (I) layer contains only $\ket{0}$ and $\ket{+}$, and conversely. Pauli gates move into the frame, as in step~6 of the simulator of Ref.~\cite{GJ26}. Mid-circuit measurements are deferred with a fresh ancilla $a$:
\begin{align}
    \MZ^{(q)} &\equiv \MZ^{(a)}\,\CX{q}{a}, \quad a\ \text{in}\ \ket{0},\\
    \MX^{(q)} &\equiv \MX^{(a)}\,\CX{a}{q}, \quad a\ \text{in}\ \ket{+},
\end{align}

and a reset becomes a new wire initialised in $\ket{0}$ or $\ket{+}$. The converse holds because $\ket{0}$ and $\ket{+}$ are preparable by CSS circuits. The one difference is that Ref.~\cite{GJ26} also allows classically controlled CNOTs, whereas ICM keeps the CNOT network fixed and places all adaptivity in $X/Z$ basis choices.

\textit{Non-CSS-ness as a resource.}---Ref.~\cite{GJ26} asks whether non-CSS-ness can be treated as a resource encoded into states, as magic is~\cite{BK05}. The ICM compiler~\cite{PPND17} gives an explicit construction. $P=\sqrt{Z}$ is a $\ket{Y}$-teleportation. $H$ becomes three $\ket{Y}$-teleportations. $T$ is an $\ket{A}$-teleportation whose probabilistic $P$ correction is converted, through selective source and destination teleportation~\cite{Fow12}, into an additional $\ket{Y}$ input and an adaptive $X/Z$ basis choice. The distillation circuits for $\ket{Y}$ and $\ket{A}$ are themselves ICM circuits. Hence every Clifford$+T$ circuit equals a CSS circuit acting on $\{\ket{0},\ket{+},\ket{Y},\ket{A}\}$ inputs. All non-CSS-ness sits in the $\ket{Y}$ inputs and all non-stabilizerness in the $\ket{A}$ inputs, so the $\ket{Y}$-count is a natural measure of non-CSS-ness, by analogy with the $T$-count. The CSS$+H$ GHZ circuit of Ref.~\cite{GJ26}, which uses non-CSS gates only in state preparation, is an instance of this structure.

\textit{What is new in Ref.~\cite{GJ26}.}---The ICM papers did not observe that the CSS core admits an all-zero reference sample and $O(1)$-per-gate sampling, nor that a local-hidden-variable obstruction prevents extending the simulator to $H$, $S$ or CZ, nor the application to atom loss. These results fit the ICM picture: in ICM form the frame simulator covers the (C) and (M) layers, the obstruction can only enter through $\ket{Y}$ or $\ket{A}$ inputs, and a lost atom deletes CNOTs from (C) without leaving the class. We suggest that CSS circuits and $\{\ket{0},\ket{+}\}$-initialised ICM circuits be regarded as the same object.

\textit{Acknowledgements.} -- Claude Opus 5.5 was used in the writing of this paper. 


\begin{thebibliography}{9}
\bibitem{GJ26} C.~Gidney and C.~Jones, \href{https://arxiv.org/abs/2611.14944}{arXiv:2611.14944} (2026).
\bibitem{PPND15} A.~Paler, I.~Polian, K.~Nemoto, and S.~J.~Devitt, in \textit{Reversible Computation}, LNCS \textbf{9138}, 139 (Springer, 2015).
\bibitem{PPND17} A.~Paler, I.~Polian, K.~Nemoto, and S.~J.~Devitt, \href{https://doi.org/10.1088/2058-9565/aa66eb}{Quantum Sci.\ Technol.\ \textbf{2}, 025003 (2017)}, \href{https://arxiv.org/abs/1509.02004}{arXiv:1509.02004}.
\bibitem{PDNP14} A.~Paler, S.~J.~Devitt, K.~Nemoto, and I.~Polian, in \textit{Proc.\ DATE} (2014), p.~1.
\bibitem{Fow12} A.~G.~Fowler, \href{https://arxiv.org/abs/1210.4626}{arXiv:1210.4626} (2012).
\bibitem{BK05} S.~Bravyi and A.~Kitaev, Phys.\ Rev.\ A \textbf{71}, 022316 (2005).
\bibitem{Gid21} C.~Gidney, Quantum \textbf{5}, 497 (2021).
\bibitem{Kis22} A.~Kissinger, \href{https://arxiv.org/abs/2204.14038}{arXiv:2204.14038} (2022).
\end{thebibliography}
\end{document}